\documentclass[%
reprint,%
aps,%
pra,%
groupedaddress,%
superscriptaddress,%
amssymb,%
amsmath,%
floatfix%
]{revtex4-2}

\usepackage{graphicx}
\usepackage[
  colorlinks=false,
  urlcolor=black,
  linkcolor=blue,
  citecolor=blue
]{hyperref}
\usepackage{comment}
\usepackage{upgreek}

\newcommand{\ngr}{n_{\rm g}}
\newcommand{\LS}{L_{\rm S}}
\newcommand{\LL}{L_{\rm L}}

\begin{document}

\title{Experimental demonstration of topological slow light waveguides\\
in valley photonic crystals}

\author{Hironobu Yoshimi}
    \email[Correspondence email address: ]{hyoshimi@iis.u-tokyo.ac.jp}
    \affiliation{Research Center for Advanced Science and Technology, The University of Tokyo,\\ 4-6-1 Komaba, Meguro-ku, Tokyo 153-8505, Japan}
    \affiliation{Institute of Industrial Science, The University of Tokyo, 4-6-1 Komaba, Meguro-ku, Tokyo 153-8505, Japan}

\author{Takuto Yamaguchi}
    \affiliation{Research Center for Advanced Science and Technology, The University of Tokyo,\\ 4-6-1 Komaba, Meguro-ku, Tokyo 153-8505, Japan}
    \affiliation{Institute of Industrial Science, The University of Tokyo, 4-6-1 Komaba, Meguro-ku, Tokyo 153-8505, Japan}

\author{Ryota Katsumi}
    \affiliation{Research Center for Advanced Science and Technology, The University of Tokyo,\\ 4-6-1 Komaba, Meguro-ku, Tokyo 153-8505, Japan}
    \affiliation{Institute of Industrial Science, The University of Tokyo, 4-6-1 Komaba, Meguro-ku, Tokyo 153-8505, Japan}

\author{Yasutomo Ota}
    \affiliation{Institute for Nano Quantum Information Electronics, The University of Tokyo,\\ 4-6-1 Komaba, Meguro-ku, Tokyo 153-8505, Japan}

\author{Yasuhiko Arakawa}
    \affiliation{Institute for Nano Quantum Information Electronics, The University of Tokyo,\\ 4-6-1 Komaba, Meguro-ku, Tokyo 153-8505, Japan}

\author{Satoshi Iwamoto}
    \email[Correspondence email address: ]{iwamoto@iis.u-tokyo.ac.jp}
    \affiliation{Research Center for Advanced Science and Technology, The University of Tokyo,\\ 4-6-1 Komaba, Meguro-ku, Tokyo 153-8505, Japan}
    \affiliation{Institute of Industrial Science, The University of Tokyo, 4-6-1 Komaba, Meguro-ku, Tokyo 153-8505, Japan}
    \affiliation{Institute for Nano Quantum Information Electronics, The University of Tokyo,\\ 4-6-1 Komaba, Meguro-ku, Tokyo 153-8505, Japan}



\begin{abstract}
We experimentally demonstrate topological slow light waveguides in valley photonic crystals (VPhCs). 
We employed a bearded interface formed between two topologically-distinct VPhCs patterned in an air-bridged silicon slab. The interface supports both topological and non-topological slow light modes below the light line.
By means of optical microscopy, we observed light propagation in the topological mode in the slow light regime with a group index $\ngr$ over $30$. Furthermore, we confirmed light transmission via the slow light mode even under the presence of sharp waveguide bends. In comparison between the topological and non-topological modes, we found that the topological mode exhibits much more efficient waveguiding than the trivial one, elucidating topological protection in the slow light regime. This work paves the way for exploring topological slow-light devices compatible with existing photonics technologies.
\end{abstract}

\maketitle

\section{Introduction}

Slow light waveguides in photonic crystals (PhCs) have been under extensive research with prospects for a wide range of applications, including photonic integrated circuits (PICs).
They are highly attractive for advanced control over optical signals and for enhancing light-matter interactions.
Indeed, they have played key roles in downsizing optical delay lines \cite{baba2008large} and modulators \cite{hinakura201964}, as well as in enhancing optical gain \cite{ek2014slow}, nonlinearity \cite{corcoran2009green}, and quantum light emission \cite{PhysRevLett.113.093603}.
On the other hand, backscattering is known to increase in the slow light waveguides, particularly under the presence of structural imperfections and/or sharp waveguide turns. For the practical use of the slow light waveguides, it is therefore crucial to mitigate backscattering \cite{o2010loss, assefa2006transmission}.

A potential solution to the issue of backscattering is the utilization of topological photonic crystals \cite{ozawa2019topological}. To date, backscattering-immune waveguiding in the edge states of photonic quantum-Hall-like and quantum-spin-Hall-like systems has been demonstrated \cite{haldane2008possible, wang2009observation, bahari2017nonreciprocal, hafezi2011robust, khanikaev2013photonic, chen2014experimental, wu2015scheme, barik2018topological, Bandreseaar4005, PhysRevLett.125.013903}. However, the realization of topological slow light modes is not straightforward, since topological edge states by nature exhibit linear dispersions and behave as fast light. Nevertheless, there are several proposals for realizing topological slow light edge states \cite{gangaraj2018topological, chen2019switchable, chen2015manipulating, guglielmon2019broadband, yoshimi2020slow, chen2020broadband, PhysRevLett.126.027403, pujol2021slow}. Unfortunately, most of them require complex structures or materials and their realization in the optical regime is difficult. Indeed, all the experimental demonstrations to date have been conducted in the microwave frequency bands \cite{yang2013experimental, chang2020direct}.

Recently, we have reported a way to realize topological slow light in the optical wavelength bands \cite{yoshimi2020slow}. We found that bearded interfaces formed in valley photonic crystals (VPhCs) \cite{shalaev2019robust, he2019silicon, yamaguchi2019gaas, ma2019topological, JalaliMehrabad:20, noh2020experimental, zeng2020electrically, arora2021direct} can support topological slow light modes. The VPhC waveguide does not require extraordinary materials or structures, and can be implemented in normal dielectrics. Therefore, the VPhC structure could provide an alternative route to realize topological slow light waveguides compatible with existing PIC technologies.

In this paper, for the first time, we experimentally demonstrate topological slow light waveguides at telecommunication wavelengths with large group indices ($\ngr$s) over 30. We fabricated straight and Z-shaped topological waveguides formed at bearded interfaces in air-bridged VPhCs patterned in a silicon slab. Utilizing optical microscopy, we measured transmission spectra of the fabricated devices and observed waveguiding in the topological slow light mode even under the presence of sharp turns. Meanwhile, we found largely-suppressed light transmission in a trivial waveguide mode co-existing in the same VPhC interface. The contrast between the two results highlights the significance of topological protection in the propagation of slow light. Our work paves the way for implementing topological slow light waveguides in PIC platforms and thereby for augmenting PIC functionality and enhancing light-matter interactions therein.

\begin{figure*}[tb]
\centering\includegraphics[width=0.7\linewidth]{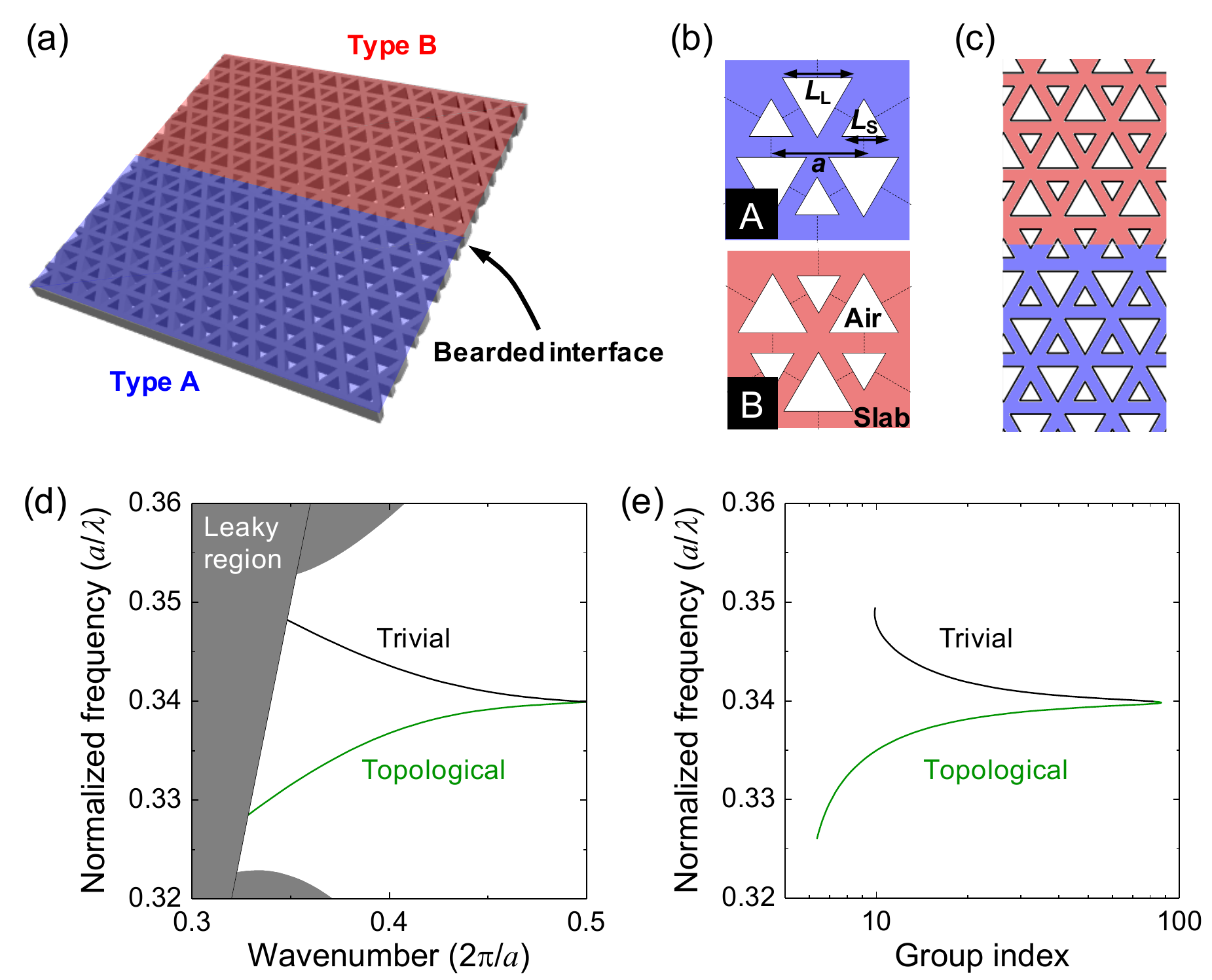}
\caption{
(a) Schematic illustration of the investigated VPhC waveguide.
(b) Schematic of the unit cells of the VPhCs.
(c) Top view of the bearded interface formed between two topologically distinct VPhCs.
(d) Dispersion curves for the topological and trivial in-gap states formed at the bearded interface in Fig. 1 (c).
(e) Calculated group indices for the in-gap states.
}
\end{figure*}

\section{Device structure}
Figure 1(a) shows a schematic illustration of the slab-type VPhC waveguide investigated in this paper.
Here, the device design is based on our previous report \cite{yoshimi2020slow}.
The waveguide is constituted of an interface between two topologically distinct VPhCs, type A (colored in blue) and type B (red).
Schematics of the unit cells of the VPhCs are shown in Fig. 1(b).
Two kinds of equilateral triangular air holes are contained in a unit cell and are arranged in a honeycomb lattice with a period $a$. 
The side lengths of the triangular air holes are denoted as $\LL$ and $\LS$, respectively.
When $\LL=\LS$, the system possesses $C_{6v}$ point group symmetry and supports a symmetry-protected Dirac cone between the first and the second lowest frequency bands at K and K' points.
On the other hand, when $\LL \neq \LS$, spatial inversion symmetry is broken in this system, resulting in the formation of a topological bandgap between the two bands. 
The band topology of the gapped band can be characterized by valley Chern numbers, which differ between the K and K' valleys in a single particular band and also between the type A and B VPhCs when evaluated at the same valley. For an interface consisted of the two topologically-distinct VPhCs, a one-dimensional topological edge state is expected to emerge. This valley kink state has been shown to function as an optical waveguide and to exhibit robust light waveguiding even under the presence of sharp waveguide bends \cite{shalaev2019robust}. In addition, it was recently reported that the propagation loss, mainly due to structural imperfections, may be reduced in the VPhC waveguides \cite{arora2021direct, PhysRevLett.126.027403}.

\begin{figure*}[tb]
\centering\includegraphics[width=0.9\linewidth]{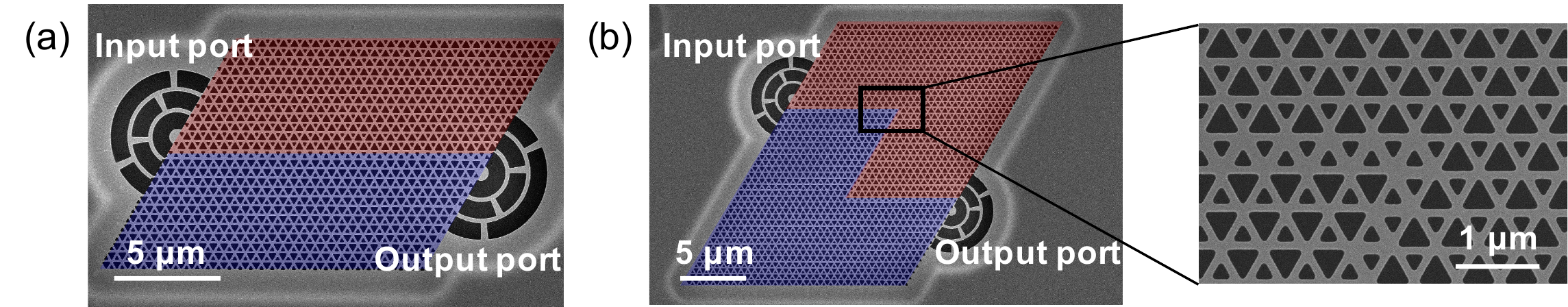}
\caption{
SEM images of (a) straight and (b) Z-shaped waveguides.
The inset in (b) shows a magnified image of a sharp bend of the waveguide.
}
\end{figure*}

Figure 1(c) shows a bearded interface formed between the two VPhCs with $\LL=1.3a/\sqrt{3}$ and $\LS=0.7a/\sqrt{3}$. Here, the smaller triangular air holes are confronted at the interface.
Figure 1(d) shows simulated dispersion curves for the edge states formed at the bearded interface with a lattice constant $a = 530 \ {\rm nm}$ and a slab thickness $d = 200 \ {\rm nm}$. The simulations were performed for the TE-like field mode utilizing the three-dimensional plane wave expansion method. The refractive index of the slab was set to 3.4.
Two guided modes are found in the gap: the lower frequency curve stems from a topological edge state, while the higher frequency curve a trivial one. The two guided modes are degenerated at the Brillouin zone (BZ) edge, which is forced by the glide plane symmetry across the interface \cite{mock2010space}. This forced degeneracy leads to bending the topological band when lowering the frequency of the trivial band by controlling $\LL$ and $\LS$. For a set of optimized air hole sizes, a single-mode slow light region can be found using the three-dimensional calculaitons, as shown in Fig. 1(d). Figure 1(e) shows calculated $\ngr$s of the in-gap states below the lightline, demonstrating large $\ngr$s of up to 80 for the topological waveguide mode.  

The origins of the two in-gap modes can be understood from the fact that the bearded interface can be transformed into a zigzag interface by modifying a single row of air holes adjacent to the interface. For the case shown in Fig. 1(c), a zigzag interface can be obtained by enlarging a row of the smaller air holes confronting the interface to the size of the larger air holes with $\LL = 1.3a/\sqrt{3}$. When gradually changing the air hole size, one finds that the trivial band rapidly increases its frequency and is finally merged with the upper bulk bands. Meanwhile, the topological band remains in the gap and is adiabatically transformed to the fast-light topological mode of the zigzag interface, which confirms the consistency of the initial mode assignment. Further detailed discussion on the origins of the modes in the bearded interface can be found in our previous publication \cite{yoshimi2020slow}.

\section{Sample fabrication and optical transmission measurements}
We fabricated straight and Z-shaped topological slow light waveguides in a 200-nm-thick silicon slab with a combination of electron beam lithography and reactive ion etching.
Air-bridged structures were formed by dissolving a sacrificial ${\rm SiO_2}$ layer
underneath the silicon slab.
We set the lattice constant of the VPhC to 530 nm so that it can operate at telecommunication wavelengths.
Scanning electron microscopy (SEM) images of the fabricated samples are shown in Figs. 2(a) and (b). The inset in Fig. 2(b) shows an enlarged image of a sharp bend in the Z-shaped waveguide.
The lengths of the straight and Z-shaped waveguides are $30a$ and $45a$, respectively.
Both waveguides are terminated by semicircular grating ports for light input and output.
The positions of the grating ports are intentionally shifted from the waveguide center along with the bulk VPhC. This allows for the control of the light reflectance at the waveguide ends. For highly reflective cases, we observed clear Fabry-P\'{e}rot (FP) fringes in transmission spectra, from which the $\ngr$ spectra of the waveguide modes can be deduced \cite{notomi2001extremely}.

To characterize the guided modes, we first measured transmission and $\ngr$ spectra for the straight and Z-shaped waveguides.
We utilized a laser-driven xenon lamp with a broadband spectrum as a light source.
Input light beam is focused on a grating port by an objective lens ($\times50$, numerical aperture (N.A.): 0.65) from the top of the sample and is coupled into the waveguide.
Output light from the waveguide is collected by the same objective lens and sent to a grating spectrometer with a cooled InGaAs camera.
Figure 3(a) shows a transmission spectrum for the straight waveguide. 
As expected, the spectrum contains sharp peaks originated from the FP resonances sustained by the reflection at the waveguide ends.
From the FP fringes, $\ngr$s of the waveguide modes are calculated and plotted as blue dots in Fig. 3(b). The deduced $\ngr$s form a peak at around $\lambda = 1470\ {\rm nm}$ and rapidly decay as moving away from the peak.
In the figure, we also plot computed $\ngr$ spectra by the three-dimensional plane wave expansion method with an offset in the horizontal axis for better comparison with the experimental data.
The measured $\ngr$ spectra agree well with the computed curves of the trivial and topological waveguide modes, suggesting the experimental observation of light transmission in both two modes.

Next, we measured a transmission spectrum for the Z-shaped waveguide designed with exactly the same VPhC parameters. The obtained spectrum is shown in Fig. 3(c). In contrast to the spectrum in Fig. 3(a), an obvious difference between the two curves can be observed in the shorter wavelength region of $\lambda \leq 1470\ {\rm nm}$.
At these wavelengths, almost no light transmission was observed for the Z-shaped waveguide. In contrast, clear FP fringes were observed in the longer wavelength band of $\lambda > 1470\ {\rm nm}$.
These results are consistent with our mode assignments. The shorter wavelength band corresponds to the trivial guided mode and is thus heavily back-reflected at the bends, resulting in the strong suppression of light output from the grating port. The longer wavelength mode is topological and can efficiently travel through the bends. These behaviours are also consistent with the results from numerical computation reported in our previous work \cite{yoshimi2020slow}.
Figure 3(d) shows measured $\ngr$s for the topological mode in the Z-shaped waveguide. 
The measured $\ngr$s increase towards $\lambda = 1470\ {\rm nm}$.
The maximum observed $\ngr$ is $\sim 37$, which is approximately eight times larger than that in normal silicon wire waveguides \cite{dulkeith2006group} and elucidates a significant slow down of the guided topological mode. These results demonstrate light transmission through the topological slow light waveguide even under the presence of sharp waveguide turns.
We note that the experimental results above are highly reproducible. We compared the spectra of straight and Z-shaped waveguides fabricated spatially close to each other. The two devices are only $340\ {\rm \upmu m}$ apart. In such cases, the VPhCs are patterned almost identically and can be precisely compared. Another note is that the fringe spacing observed in Fig. 3(c) is narrower compared to that in Fig. 3(a); this occurs simply because the Z-shaped waveguide is longer than the straight one.

\begin{figure}[tb]
\centering\includegraphics[width=\linewidth]{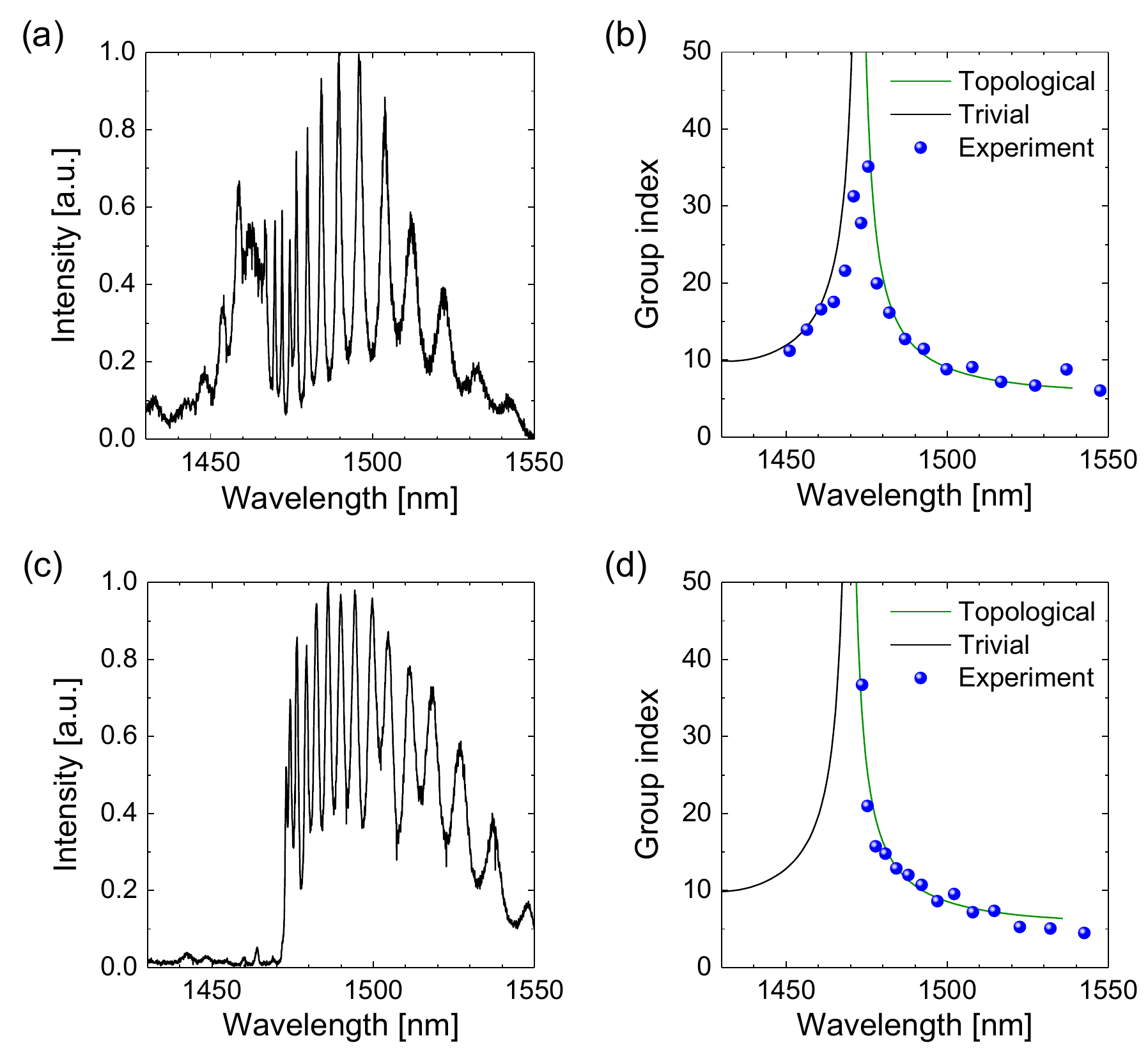}
\caption{
Measured transmission and $\ngr$ spectra for (a,b) straight and (c,d) Z-shaped waveguides.
Solid curves in (b) and (d) show calculated $\ngr$s by the three-dimensional plane wave
expansion method. The green and black curves correspond to the topological and trivial modes, respectively. In (d), measured $\ngr$s are plotted only for the topological band.
}
\end{figure}

\section{Light propagation imaging}
Next, we investigate propagation characteristics of the topological and trivial modes through sharp waveguide bends using optical microscope imaging. We examined another Z-shaped waveguide patterned with the same VPhC parameter but without shifting the position of the grating output port for more efficient light output. The investigated waveguide is located just next to the one studied in Fig. 3. Therefore, as discussed in the previous section, it is safe to estimate that the current waveguide provides almost the same wavelength dependence of $\ngr$s as those of the device in Fig. 3. 
For this experiment, we utilized a wavelength-tunable single-mode laser diode as a light source. By varying the wavelength, we can selectively excite each guided mode at a respective $\ngr$. The optical images were taken through an objective lens with a higher magnification of $\times 100$ and a higher N.A. of 0.8, using an InGaAs camera. During the imaging experiments, we fixed the position of the sample and laser spot, and set the laser input power to 400 nW, so as to extract solely the influence of input laser wavelength.

Figure 4 shows microscope images of the Z-shaped waveguide measured at various input laser wavelengths. 
In the figure, the green solid and broken lines respectively represent the outer frame and waveguide of the VPhCs.
Figures 4(a) and (b) show the results when the input laser light couples to the trivial band. For the input wavelength of $\lambda$ = 1460 nm (1450 nm), $\ngr$ of the trivial waveguide mode is estimated to be $\ngr \sim 20 \ (13)$. In either case, almost no output light was observed from the output grating port. Moreover, scattered light at the first waveguide bend was clearly detected, which became more prominent in the case of $\ngr \sim 20$. These results are consistent with the transmission measurements shown in Fig. 3, where light in the trivial band hardly propagates through sharp waveguide bends. The results here suggest that light in the trivial band is easily scattered out to free space or back-reflected at the corners.
On the other hand, Figs. 4(c)-(f) show microscope images when the laser wavelength is within the band of the topological edge state.
Input wavelengths and estimated $\ngr$s are as follows: (c) $\lambda$ = 1473.2 nm, $\ngr \sim 30$, (d) $\lambda$ = 1475.6 nm, $\ngr \sim 20$, (e) $\lambda$ = 1485 nm, $\ngr \sim 12$, and (f) $\lambda$ = 1497 nm, $\ngr \sim 8$.
In all these figures, clear output light is observed from the grating port. Moreover, scattering-free transmission is observed for lower $\ngr$s of 8 and 12. Even when increasing $\ngr$ to 30, there is only a minor increase in light scattering at the corners and a weak drop in the intensity at the output port. Indeed, the light spots faintly visible at the corners are still much weaker than those at the grating output port. These observations demonstrate robust light transport in the topological slow light waveguide even under the presence of sharp waveguide bends. The observed stark contrast between the topological and trivial modes elucidates high impact of topological protection in efficient light transport in slow light waveguides.

\begin{figure}[tb]
\centering\includegraphics[width=\linewidth]{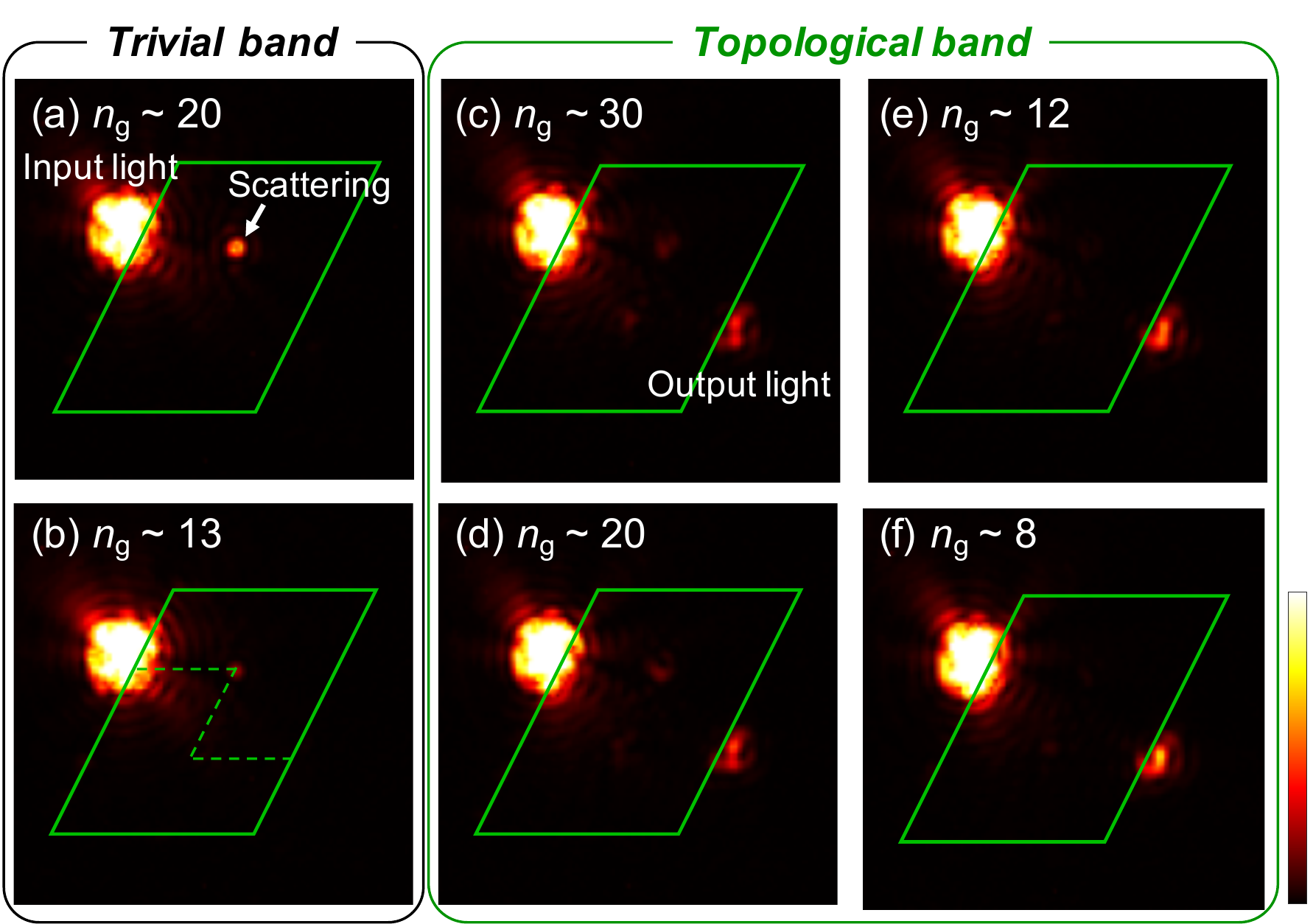}
\caption{
Near-infrared microscope images measured with laser irradiation exciting the trivial (a,b) and topological (c-f) kink states. Solid and broken green lines indicate the positions of the outer frame and the waveguide of the VPhC, respectively. The excitation wavelengths and estimated $\ngr$s in each figure are as follows: (a) $\lambda$ = 1460 nm, $\ngr \sim 20$, (b) $\lambda$ = 1450 nm, $\ngr \sim 13$, (c) $\lambda$ = 1473.2 nm, $\ngr \sim 30$, (d) $\lambda$ = 1475.6 nm, $\ngr \sim 20$, (e) $\lambda$ = 1485 nm, $\ngr \sim 12$, and (f) $\lambda$ = 1497 nm, $\ngr \sim 8$.
}
\end{figure}

\section{Summary}
In summary, we experimentally demonstrated a topological slow light waveguide operating at telecommunication wavelengths. We employed a slow-light topological kink state sustained at a bearded interface in VPhCs and fabricated the designed structures in a silicon slab. Utilizing optical microscopy, we observed light propagation through the topological slow light waveguide even under the presence of sharp waveguide bends.
These results provide important implications for the development of PICs employing topological photonics \cite{iwamoto2021recent}, active topological photonic devices \cite{ota2020active} and in the study of light-matter and nonlinear optical phenomena in topological optical modes \cite{smirnova2020nonlinear}.

\section*{Funding}

This work was supported by JSPS KAKENHI Grant-in-Aid for Specially Promoted Research (15H05700),
JSPS KAKENHI (15H05868 and 17H06138), JST CREST (JPMJCR19T1), and The Asahi Glass Foundation Research Grant Program.


\section*{Disclosures}

The authors declare no conflicts of interest.








\bibliography{sample}






\end{document}